\documentclass[12pt]{article}
\usepackage{graphicx}
\usepackage{booktabs}
\usepackage{dcolumn}
\usepackage{fullpage}
\usepackage{times}
\usepackage{amsmath}
\usepackage{amsfonts}
\usepackage{amssymb}
\usepackage{setspace}
\usepackage[numbers]{natbib}
\usepackage{rotating}
\usepackage{xcolor}
\usepackage{ulem}
\usepackage{hyperref}
\def\notationcolor{black} 

\newcommand{\notation}[2]{\newcommand{#1}{\textcolor{\notationcolor}{\ensuremath{#2}}}}

\notation{\tauhi}{\tau^*}  
\notation{\taulow}{\tau}   
\notation{\usermax}{C_u} 
\notation{\numrows}{d} 
\notation{\numcols}{m} 
\notation{\M}{M} 
\notation{\X}{X} 
\notation{\Xminusj}{X_{-j}} 
\notation{\Null}{\emptyset} 
\notation{\aplus}{a_+} 
\notation{\aequals}{a_=} 
\notation{\aminus}{a_-} 
\notation{\Fj}{F_j} 
\notation{\muzero}{\mu_o} 

\usepackage{caption}
\usepackage{subcaption}

\usepackage[ruled,vlined, linesnumbered]{algorithm2e}

\newtheorem{theorem}{Theorem}[section]
\newtheorem{corollary}{Corollary}[theorem]
\newtheorem{lemma}[theorem]{Lemma}
\newtheorem{observation}[theorem]{Observation}
\newtheorem{definition}{Definition}[section]

\DeclareMathOperator{\Prob}{{\mathbb{P}}}

\usepackage{multirow} 

\usepackage{authblk}

\begin{document}

\title{Exact Privacy Analysis of the Gaussian Sparse Histogram Mechanism \singlespacing}
\author[1]{Brian Karrer}
\affil[1]{Core Data Science, Meta}
\author[2]{Daniel Kifer}
\affil[2]{Pennsylvania State University, University Park, PA}
\author[1]{Arjun Wilkins}
\author[2]{Danfeng Zhang}
\date{}
\maketitle

\begin{abstract}
Sparse histogram methods can be useful for returning differentially private counts of items in large or infinite histograms, large group-by queries, and more generally, releasing a set of statistics with sufficient item counts.  We consider the Gaussian version of the sparse histogram mechanism and study the exact $\epsilon,\delta$ differential privacy guarantees satisfied by this mechanism.  We compare these exact $\epsilon,\delta$ parameters to the simpler overestimates used in prior work to quantify the impact of their looser privacy bounds.
\end{abstract}

\section{Introduction}\label{sec:intro}
Consider a dataset with a very large, or possibly infinite domain, such as a dataset of user interactions with a URL --  every time a user from country $C$ performs an action $A$ (e.g., share, like) on a url $U$, the record $(U, C, A)$ is added to the data. It is natural to ask group-by queries, such as
\begin{verbatim}SELECT COUNT(*) FROM table WHERE Action=``share'' 
    GROUP BY URL, COUNTRY
\end{verbatim}
which counts the number of ``shares'' a URL has in each country.
Answering this type of query under pure differential privacy would essentially require enumerating every \emph{possible} URL and country combination -- this is called a \emph{Cartesian expansion} (of the grouping columns URL and Country). Note that under pure differential privacy, a Cartesian expansion also includes combinations that have zero counts in the dataset, since adding or removing an individual in the dataset might change those counts. Clearly, computing a Cartesian expansion is infeasible for extremely large  (or, as in this case, infinite) domains.
\footnote{Another example where the domain is too large for this to be practical is returning counts of appearances of n-grams ($n$ consecutive words) from a text corpus.}

In such cases, one settles for approximate differential privacy and thresholding schemes \cite{korolova09,Balcer_Vadhan_2019,wilson_et_al,google_add_deltas, Gotz2012, Bun2016}: one first filters out items whose true counts are 0,  adds noise to the remaining items, then returns the noisy counts for items  whose noisy counts meet or exceed some threshold \tauhi.  This parameter $\tauhi$ should be set high enough so that even if noise were added to a count whose true value was 0, the noisy value will, with overwhelming probability, still be less than $\tauhi$. This mechanism is known in the literature as a \emph{sparse histogram} or stability histogram mechanism \cite{Balcer_Vadhan_2019}.

Due to the vague nature of privacy laws today, a slight generalization of this approach may also need to be considered: one first filters out items whose true counts are less  than \taulow, adds noise to the remaining items, and then returns those noisy counts for items whose noisy counts meet or exceed a second threshold \tauhi. The reason for this generalization is that some laws could be interpreted as prohibiting statistics computed from $k$ or fewer people (for some $k$). Thus this generalization combines the ``required'' cell suppression and differential privacy (which provides the mathematical privacy guarantees).  This generalization can also be useful when a relevant scale for sparsity is some non-zero value. 

In this paper, we study this double thresholding mechanism and derive an exact $\epsilon,\delta$ curve for the case where the noise used is Gaussian. The reason for this specific emphasis on the Gaussian is that many end-users are more comfortable with this distribution for their subsequent statistical analyses. We compare the exact $\epsilon,\delta$ curve to the approach of Wilson et al. \cite{wilson_et_al,google_add_deltas}, which provides an over-approximation of the privacy parameters (originally they derived their results for Laplace noise and $\taulow=1$, thus filtering out exactly those cells with 0 counts; later they extended the work to Gaussian noise, but still with $\taulow=1$). One goal of this paper is to quantify the impact of this over-approximation and to identify when a more exact privacy loss accounting is necessary.

Our contributions are the following:
\begin{itemize}
\item We derive the exact $\epsilon,\delta$ curve for the Gaussian noise-based sparse histogram mechanism. In the database setting, this is equivalent to a group-by query that returns group sizes along with other aggregations for each group, but filters out small groups.
\item We provide a case study that allows us to analyze the impact of the conservative $\epsilon,\delta$ calculations used in prior work.
\end{itemize}

This paper is organized as follows. In Section~\ref{sec:background}, we present relevant background material on differential privacy and the Gaussian mechanism.  Then in Section~\ref{sec:gshm} we review the Gaussian sparse histogram mechanism.  Notation introduced in these sections is summarized in Table~\ref{tab:notation}.  Then in Section~\ref{sec:related_work}, we present related work on sparse histogram mechanisms, prior to deriving our exact privacy analysis of the Gaussian sparse histogram mechanism in Section~\ref{sec:privacy_analysis}.  In Section~\ref{sec:alt_case_study}, we compare our results against privacy accounting approaches in prior work on a case study and then in Section~\ref{sec:conclusion} we present conclusions.


\begin{table}
\begin{tabular}{|c||c||l|}\hline
\multirow{2}{*}{Data} & \X & Dataset \\\cline{2-3} 
& \Xminusj & Dataset $\X$ with user $j$ removed \\\hline
\multirow{5}{*}{Mechanism} & \numrows & Number of potential rows in output ($1 \leq \numrows \leq \infty$) \\\cline{2-3}
& \numcols & Number of columns per output row ($\numcols \geq 1)$ \\\cline{2-3}
& \Null & Null output value; simply not returned \\\cline{2-3}
& \M$(\X)$ & random mechanism applied to dataset $\X$ returning $\{\Null, \mathbb{R}\}^{d \times m}$ \\\cline{2-3}
& \M$(\X)_i$ & $i$th output row of mechanism $\in \{\Null, \mathbb{R}\}^m$ \\\hline
\multirow{4}{*}{Parameters} & \taulow & Low non-negative threshold\\\cline{2-3}
& \tauhi & High non-negative threshold ($\tauhi > \taulow$)\\\cline{2-3}
& $\sigma$ & Standard deviation of noise for user count column \\\cline{2-3}
& $\Sigma$ & Covariance matrix of noise for remaining $\numcols-1$ columns \\\hline
\multirow{7}{*}{Privacy analysis}
& $(\epsilon, \delta)$ & Approximate differential privacy parameters \\\cline{2-3}
& \usermax & Maximum number of rows a user can affect \\\cline{2-3}
& \aplus & Number of rows affected by user $j$ with user count above \taulow \\\cline{2-3}
& \aequals & Number of rows affected by  user $j$ with user count equal to \taulow \\\cline{2-3}
& \aminus & Number of rows affected by user $j$ with user count below \taulow \\\cline{2-3}
& \Fj & A row affected by user $j$ with user count equal to $\taulow$ is not $\Null^{m}$ \\\cline{2-3}
& \muzero & $\mu$ contribution from remaining columns when $\numcols > 1$ \\\hline
\end{tabular}
\caption{Table of notation for Gaussian sparse histogram mechanism}\label{tab:notation}
\end{table}

\section{Background}\label{sec:background}
Differential privacy  is an emerging gold standard for settings where a data release mechanism must process private data and produce publicly shareable information while each individual's privacy is protected. Differential privacy is a set of restrictions on the behavior of the data release mechanism.  Roughly speaking, a privacy mechanism   is differentially private if the probability distribution of the output of the mechanism is fairly insensitive  to any individual's contribution to the input dataset (the probability is with respect to the randomness in the mechanism, not the randomness in the data).  
Formally,
\begin{definition}[Neighbors]
Two datasets $\X$ and $\X'$ are neighbors if one can be obtained from the other by adding records from one individual (an individual can contribute multiple records to a dataset).
\end{definition}

\begin{definition}[Approximate differential privacy \cite{dwork06Calibrating,dworkKMM06:ourdata}]\label{def:adp}
Let $\epsilon \geq 0$ and $\delta \in [0,1]$.  A randomized mechanism $\M$ satisfies $(\epsilon, \delta)$-DP if for every  pair of neighbors $\X$ and $\X'$ , and every output set $S$, we have
\begin{align} 
\Prob(\M(\X) \in S) \leq e^{\epsilon}\Prob(\M(\X') \in S) + \delta.
\end{align}
\end{definition}

The $\epsilon,\delta$ parameters of a mechanism $\M$ are typically derived with the help of a mathematical construct called the \emph{privacy loss random variable} (PLRV), which is defined as follows.

\begin{definition}[PLRV]
\label{def:plrv}
For a randomized mechanism $M$, two neighboring inputs $\X$ and $\X'$, and an output $\omega$, let $l_{\M, \X, \X'}(\omega)= \log \frac{\Prob(\M(\X) = \omega)}{\Prob(\M(\X')=\omega)}$.  Then $L_{\M, \X, \X'}$ is the \textbf{privacy loss random variable} defined as the distribution of $l_{\M, \X, \X'}(\omega)$ when $\omega$ is sampled from the distribution $\Prob(\M(\X))$.
\end{definition}

Approximate differential privacy can then be written in terms of PLRV's.

\begin{theorem}[Theorem 5 from ~\cite{balle_wang}]
A randomized mechanism $\M$ is $(\epsilon, \delta)$-DP if and only if for every pair of neighboring datasets $\X$ and $\X'$ the following holds for the associated PLRV's:
\begin{align} 
\Prob(L_{\M, \X, \X'} \geq \epsilon) - e^{\epsilon} \Prob(L_{\M, \X', \X} \leq -\epsilon) \leq \delta.
\label{eq:plrv_and_dp}
\end{align}
\label{th:plrv_and_dp}
\end{theorem}

One important noise distribution for mechanisms satisfying approximate differential privacy is the Gaussian distribution, which leads to the concept of a Gaussian mechanism:

\begin{definition}
Let $f$ be a function (known as a ``query'') whose input is a database and output is a vector of real numbers in $\mathbb{R}^{\numcols}$.  The Gaussian mechanism with covariance $\Sigma$ is the mechanism that outputs $f(\X) + Z$, where $Z \in \mathbb{R}^{\numcols}$ is drawn from $\mathbb{N}(0, \Sigma)$.
\end{definition}

The exact $\epsilon,\delta$ parameters for this mechanism can be computed using the following theorem:
\begin{theorem}[Analytic Gaussian mechanism privacy  \cite{balle_wang,dong_roth_su,XiaoDWZK21}]\label{thm:agm}
The Gaussian mechanism is $(\epsilon, \delta)$-DP if and only if
\begin{align} 
&\Phi\left(\frac{\mu}{2} - \frac{\epsilon}{\mu}\right) - e^{\epsilon} \Phi\left(-\frac{\mu}{2} - \frac{\epsilon}{\mu}\right) \leq \delta
\label{eq:analytic_gaussian}
\end{align}
where $\Phi$ is the CDF of the standard normal distribution and $\mu$ is
\begin{align}
\mu &= \max_{\text{neighboring } \X, \X'} \sqrt{ \left(f(\X) - f(\X')\right)^T\Sigma^{-1}\left(f(\X)-f(\X')\right)}.
\label{eq:mu}
\end{align}
In particular, when $\Sigma$ is a diagonal matrix with diagonals $\sigma^2_1,\sigma^2_2,\dots$ then $$\mu=\max_{\text{neighboring} \X, \X'} \sqrt{\sum_i \left(f(\X)_i - f(\X')_i\right)^2/\sigma^2_i}.$$
Furthermore, the quantity in Equation \ref{eq:analytic_gaussian} is a monotonically increasing function of $\mu$.
\label{th:analytic_gaussian}
\end{theorem}

We note that privacy interpretations require $\epsilon\geq 0$ and $\delta\geq 0$, however
we also note that Definition \ref{def:adp} is still mathematically well-defined even when $\epsilon<0$ or $\delta<0$. Furthermore,  the proofs of Theorem \ref{th:analytic_gaussian} \cite{balle_wang,dong_roth_su,XiaoDWZK21} also make no assumptions on $\epsilon$ and do not require Equation \ref{eq:analytic_gaussian} to be positive. This observation turns out to be useful for the results of this paper.
\begin{observation}
The explicit PLRV expressions in Equation~\ref{eq:analytic_gaussian} for the Gaussian mechanism hold for any value of $\epsilon$ including $\epsilon < 0$.
\label{lemma:no_positive_requirement}
\end{observation}

\section{Gaussian sparse histogram mechanism}\label{sec:gshm}
In the introduction, we briefly described a generalization of the sparse histogram mechanism that avoids Cartesian expansion through a combination of two thresholds: $\taulow$ for cell-suppression and $\tauhi$ for noisy thresholding with Gaussian noise.  Here we introduce this Gaussian sparse histogram mechanism in detail.  Let $\X=\{x_i\}_{i=1}^{n}$ be a dataset of $n$ records, $x_i \in \mathbb{X}$.

We are interested in queries that partition the records in $\X$ into $\numrows$ groups. For every group, $\numcols$ statistics are computed, one of which is the number of records in the group. The statistics should only be reported for groups that are large enough, having at least $\taulow$ records. For simplicity of presentation, we consider the setting where a user can contribute to at most $\usermax$ records, with each one belonging to a different group. Thus each user affects the counts in at most $\usermax$ groups by at most 1 per group. Note that this is the same setting as studied by Wilson et al. \cite{wilson_et_al,google_add_deltas}.

This is a natural setting for group-by queries. 
Consider the URL example from Section \ref{sec:intro}. 
Each record has the form (\emph{user id}, \emph{URL}, \emph{country}, \emph{view}, \emph{like}, \emph{share}); it records which actions (view, like, share) a user from the country has ever performed on the URL. Here view, like, and share are Boolean (0/1-valued) attributes. Note that a user can only share or like a URL if viewed. Each user is limited to $\usermax$ records and we are interested in group-by queries such as:
\begin{verbatim}
SELECT COUNT(*) AS cnt, 
   SUM(likes) AS likes,
   SUM(shares) AS shares
FROM user_url_country_table 
GROUP BY url, country
HAVING cnt >= tau
\end{verbatim}
Note that each user contributes a count of 1 to each of at most $\usermax$ groups and the number of views is actually the number of records in each group.

Here $\numrows$, the number of groups is equal to the number of countries times the number of URLs (which may be infinite) and the number of aggregates per group, $\numcols$, is 3. This output can be represented as a table with at most $\numrows$ rows and $\numcols$ columns.

The Gaussian sparse histogram mechanism introduces a second threshold $\tauhi>\taulow$ and release noisy group statistics for all groups whose noisy counts are greater than or equal to $\tauhi$. In this running example, it would look like the following SQL query:

\begin{verbatim}
SELECT noisy_cnt, noisy_likes, noisy_shares FROM (
    SELECT 
      COUNT(*) AS user_cnt, 
      COUNT(*) + GaussianNoise_1 AS noisy_cnt,
      SUM(likes) + GaussianNoise_2 AS noisy_likes,
      SUM(shares) + GaussianNoise_3 AS noisy_shares
    FROM user_url_country_table 
    GROUP BY url, country
    HAVING user_cnt >= tau AND noisy_cnt >= tau_star
)
\end{verbatim}

Formally, the mechanism is denoted as $\M$ and its goal is to  privately answer a group-by aggregation query that groups the records of $\X$ into $\numrows$ groups and computes $\numcols$ noisy aggregates for each group.
One of the aggregates must be count, and the rest can be arbitrary (as long as their sensitivity is known).
Its pseudocode is shown in Algorithm \ref{algo:gshm}.

\begin{algorithm}[]
\caption{\textbf{Gaussian sparse histogram mechanism (GSHM)}}
\label{algo:gshm}
\KwIn{User-group aggregated dataset $\X=\{x_i\}_{i=1}^{n}$ and groups $\{G_1, ... G_d\}$ where each user has at most one record in $\X$ per group and affects at most $\usermax$ groups. 
Optional aggregation function $A$.
Parameters $\taulow$, $\tauhi$, $\sigma$, and $\Sigma$ if $A$ is provided.}
\KwOut{Sparse dictionary mapping group index to noisy aggregates}

$SparseGroupAggregates = \{\}$

\For{each nonempty group $G_i$}{

Sample $v \sim \mathbb{N}(0, \sigma^{2})$

$C = $ (number of records from $\X$ in group $G_i$)

\uIf{$C \geq \taulow$ and $C + v \geq \tauhi$}{

Sample $m \sim \mathbb{N}(0, \Sigma)$

$SparseGroupAggregates[i] = [C + v, A(X, G_i) + m]$

}

}

\Return $SparseGroupAggregates$

\end{algorithm}

Thus, conceptually, its output can be organized as a $\numrows\times\numcols$ matrix, where each entry comes from the domain $\{\Null, \mathbb{R}\}^{\numrows \times \numcols}$. Groups that are filtered out are represented as rows full of $\emptyset$. We let $\M(\X)_i \in \{\Null, \mathbb{R}\}^{\numcols}$ denote the $i$th row of the output (i.e., aggregations over the $i$th group of records).

When analyzing the privacy properties of $\M$, we will make use of the following notation.
Let $G_i \subseteq \mathbb{X}$ be the set of possible records corresponding to the $i$th group (groups are disjoint).  Without loss of generality, we assume the dataset $\X$ has been aggregated per-user and per-group such that each user has at most one record per group and each record is $(userID, groupID, otherInfo)$.  Let $C(\X, G_i)$ be the number of records from $\X$ in group $G_i$ (i.e., the \underline{c}ount) and let  $A$ be an optional aggregation function that returns a vector of $\numcols-1$ real values for a group (i.e., $A(\X, G_i)\in \mathbb{R}^{\numcols-1}$).
Examples of such an $A$ include the number of shares and likes in a group, but in general, could be arbitrary as long as its privacy impact $\mu$ (see Theorem \ref{thm:agm}), after adding $N(0,\Sigma)$ noise  can be calculated.  With this notation, row $i$ in the output of the Gaussian sparse histogram mechanism $\M$ can be written as
\[
\M(\X)_i = 
\begin{cases}
\Null^{\numcols} & \text{if } C(\X, G_i) < \taulow \text{ or }C(\X, G_i) + v_i < \tauhi \\
\{C(\X, G_i) + v_i, A(\X, G_i) + m_i\} & \text{otherwise}
\end{cases}
\]
where $v_i$ is univariate Gaussian noise with standard deviation $\sigma$ and $m_i$ is multivariate Gaussian noise with covariance matrix $\Sigma \in \mathbb{R}^{(\numcols-1) \times (\numcols-1)}$. The noise of $\M$ is independent across all rows $i$.
If $\numcols=1$ (i.e., the only aggregation is the count), then there is no $A(\X, G_i)$ part.

We summarize relevant notation introduced so far for the Gaussian sparse histogram mechanism in Table~\ref{tab:notation} within the data, mechanism, and parameters sections.  Additional terms defined for our upcoming privacy analysis are also listed there for convenience.

\section{Related work}\label{sec:related_work}
Sparse histogram methods using Laplace noise were proposed for releasing click and search logs in~\cite{korolova09, Gotz2012} and also analyzed in~\cite{Bun2016}.  An overview of such sparse histogram approaches, including error bounds  can be found in~\cite{Balcer_Vadhan_2019}.  To our knowledge, this past research has not specifically considered Gaussian noise.

Accounting for unknown or large domains has also been considered for the related problem of top-$k$ selection, in \cite{Durfee2019}.  Like sparse histograms, this research involves a data-dependent pruning of outputs, in this case returning at most $k$ items whose noisy counts are large compared to the noisy $k^\prime$th element (with $k^\prime>k$). It is worth noting that top-$k$ algorithms return the identities of large items but not an estimate of their counts.

Returning to group-by queries,
instead of thresholding small groups first to achieve sparsity and then adding noise, one could consider a postprocessing approach that first adds noise to each group and then removes cells with with noisy counts less than a threshold $\tauhi$. This approach would satisfy pure differential privacy and could even be implemented efficiently (without enumerating all groups in the Cartesian expansion) when Laplace noise is used \cite{Cormode2012DifferentiallyPS}.
However, to achieve a desired level of sparsity, the threshold $\tauhi$ has to increase with the (logarithm of the) size of the Cartesian expansion, indicating dataset utility could be reduced by post-processing approaches in high-dimensional or infinite settings. 

Sparse histogram methods with Laplace noise were applied by Wilson et al. \cite{wilson_et_al} as part of  a differentially private SQL system, where avoiding Cartesian expansion was helpful for implementing group-by operations efficiently.  They additionally propose composing the count part of the query with other aggregations using the Laplace mechanism.  They extend the previous Laplace approaches of~\cite{korolova09, Gotz2012} to return multiple aggregations for each group, i.e. $\numcols > 1$.  In a later unpublished technical report \cite{google_add_deltas}, they derived an $(\epsilon, \delta)$-DP guarantee for  sparse histogram mechanisms with a wide range of noise distributions (including Gaussian noise) for the single count output ($\numcols=1$).\footnote{This derivation also includes an extension to thresholding on non-count columns with bounded positive contributions.  We do not consider this non-count extension here, but believe our results would extend to this setting.} Using the notation of our paper, their main results on the $(\epsilon,\delta)$ privacy parameters can be expressed as follows (we refer to their technique as ``add the deltas''):
\begin{theorem}[Add the deltas]
Let $\usermax$ be the maximum number of rows affected by a user.  Algorithm~\ref{algo:gshm} with $\taulow = 1$, $\tauhi$, $\numrows$, $\numcols=1$, and $\sigma$, satisfies $(\epsilon, \delta_{\text{Gaussian}} + \delta_{\text{infinite}})$-DP where
\begin{align}
\delta_{\text{Gaussian}} &= \Phi\left(\frac{\sqrt{\usermax}}{2\sigma} - \frac{\epsilon \sigma}{\sqrt{\usermax}}\right) - e^{\epsilon} \Phi\left(-\frac{\sqrt{\usermax}}{2\sigma} - \frac{\epsilon \sigma}{\sqrt{\usermax}}\right) \nonumber\\
\delta_{\text{infinite}} &= 1 - \Phi\left(\frac{\tauhi-1}{\sigma}\right)^{\usermax}.
\label{eq:m1_deltas}
\end{align}
\label{th:add_the_deltas}
\end{theorem}
We recognize $\delta_{\text{Gaussian}}$ from the Gaussian mechanism in Theorem~\ref{thm:agm} (where $\mu = \sqrt{\usermax}/\sigma$), plus another contribution  due to thresholding, $\delta_{\text{infinite}}$.  We refer to the contribution from thresholding as $\delta_{\text{infinite}}$ because it corresponds to the worst-case probability of infinite privacy loss under the mechanism.  In particular, the privacy loss random variable $l_{\M, \X, \X'}$ is infinite when $\M(\X)$ returns a row that cannot be returned by $\M(\X')$ due to the deterministic $\taulow$ threshold.  As we shall see, the worst-case probability of at least one such row being returned under the mechanism for two neighboring datasets $\X$ and $\X'$ is given by this expression for $\delta_{\text{infinite}}$.

The result in Theorem~\ref{th:add_the_deltas} is overly conservative. Next, we will derive our exact result and compare it against this theorem. In addition to tighter accounting, our result is also applicable to arbitrary $\taulow\geq 1$ and $\numcols \geq 1$ for Gaussian noise.

\section{Privacy analysis}\label{sec:privacy_analysis}
In this section, we analyze the privacy guarantees provided by the Gaussian sparse histogram mechanism. Recall that, as in prior work, ~\cite{wilson_et_al,google_add_deltas}, each user contributes at most 1 record to up to $\usermax$ groups. 
We also define 
\begin{align}
    \mu_o^2 = \max_{i \text{ and neighboring } \X, \X'} (A(\X, G_i)-A(\X', G_i))\Sigma^{-1}(A(\X, G_i)-A(\X', G_i)).
\end{align}
which summarizes the contribution of $A$ and noise covariance $\Sigma$ to the $\epsilon,\delta$ curve in Theorem \ref{thm:agm} (and indirectly in Theorem~\ref{th:plrv_and_dp}).

We use privacy loss random variables (Definition~\ref{def:plrv}) and Theorem~\ref{th:plrv_and_dp} to obtain the exact $\epsilon,\delta$ curve for the Gaussian sparse histogram mechanism. So we begin by setting up the relevant privacy loss random variables.  Without loss of generality, the target person $j$ we consider for analyzing DP properties is the first person and the output rows she affects are the first $C\leq \usermax$ rows.
Among those $C$ rows, we use $\aplus$, $\aequals$, $\aminus$ to denote the number of rows whose true user count (when $\X$ is the input) is above, equal to, below the threshold $\taulow$ respectively. Note that $\aplus+\aequals+\aminus=C\leq \usermax$.

There are two types of privacy loss random variables (dependence on $\M, \X, \Xminusj, \aplus, \aminus, \aequals$ omitted from the notation) for our mechanism $\M$:
\begin{itemize}
    \item $L_+$ is defined as the distribution of $\log\frac{\Prob(\M(\X)=\omega)}{\Prob(\M(\Xminusj)=\omega)}$, where $\omega$ is sampled from the distribution $\Prob(\M(\X))$ and $\Xminusj$ is the dataset with user $j$ removed.
    \item $L_-$ is defined as the distribution of $\log\frac{\Prob(\M(\Xminusj)=\omega)}{\Prob(\M(\X)=\omega)}$, where $\omega$ is sampled from the distribution $\Prob(\M(\Xminusj))$.
\end{itemize}
The rows where the target person does not contribute (i.e. rows after row $C$) do not affect the privacy loss random variable. The same is true with the rows where the count is below the threshold (when $\X$ is the input). Therefore the privacy loss random variables are only affected by $\aplus$ and $\aequals$ and the condition that $\aplus + \aequals\leq \usermax$.

Because each output row is independent, we can write 
\begin{align}
L_{+} &= L^{+}_{+} + L^{=}_{+} \nonumber\\
L_{-} &= L^{+}_{-} + L^{=}_{-}.
\end{align}
where $L^{+}_{+}$ is the PLRV over the $\aplus$ rows (rows containing user $j$ and above the threshold) and $L^{=}_{+}$ is the PLRV over the $\aequals$ rows (containing user $j$ and at the threshold $\taulow$), similarly for $L^{+}_{-}$ (PLRV for the same  $\aplus$ rows, but now user $j$ is removed) and $L^{=}_{-}$. 

For our mechanism to be $(\epsilon, \delta)$-DP per Theorem~~\ref{th:plrv_and_dp}, we require the following two expressions hold for any values of $\aplus + \aequals \leq \usermax$.
\begin{align}
&\Prob(L_{+} \geq \epsilon) - e^{\epsilon} \Prob(L_{-} \leq -\epsilon) \leq \delta \nonumber\\
&\Prob(L_{-} \geq \epsilon) - e^{\epsilon} \Prob(L_{+} \leq -\epsilon) \leq \delta.
\label{eq:plrv_expression_for_dp}
\end{align}
for every $\X$ and $j$.  The first expression corresponds to $\X$ containing $j$ and $\X'$ not containing $j$, and vice versa.  

Next, we evaluate PLRVs under two cases (depending on whether $\aplus = 0$ or not), where all proofs are available in the Appendix.
\begin{lemma}[Case \aplus $= 0$]
If \aplus$=0$, Equation~\ref{eq:plrv_expression_for_dp} is satisfied when
\begin{align} 
1 - \Phi\left(\frac{\tau^{*} - \tau}{\sigma}\right)^{C_u} \leq \delta 
\end{align}
\label{lemma:aequals}
\end{lemma}

\begin{lemma}[Case \aplus $> 0$]
If \aplus $>0$, Equation~\ref{eq:plrv_expression_for_dp} is satisfied when the following two conditions are satisfied:
\begin{align} 
&\max_{a_{+} + a_{=} \leq C_u, a_{+} > 0} 1 - \Phi\left(\frac{\tau^{*} - \tau}{\sigma}\right)^{a_{=}} + \Phi\left(\frac{\tau^{*} - \tau}{\sigma}\right)^{a_{=}} [ \Prob(L^{+}_{+} \geq \epsilon_2) - e^{\epsilon_2} \Prob(L^{+}_{-} \leq -\epsilon_2)] \leq \delta, \nonumber\\
&\max_{a_{+} + a_{=} \leq C_u, a_{+} > 0} \Prob(L^{+}_{-} \geq \epsilon_3) - e^{\epsilon_3} \Prob(L^{+}_{+} \leq -\epsilon_3) \leq \delta,
\end{align}
where
\begin{align}
\epsilon_2(a_{=}) = \epsilon - a_{=} \log \Phi\left(\frac{\tau^{*} - \tau}{\sigma}\right), && \epsilon_3(a_{=}) = \epsilon + a_{=} \log \Phi\left(\frac{\tau^{*} - \tau}{\sigma}\right). \nonumber
\end{align}
\label{lemma:aplus}
\end{lemma}

To simplify Lemma~\ref{lemma:aplus} further, we work out the remaining PLRV terms that correspond to rows above threshold $\taulow$.  For these $\aplus$ rows, the Gaussian sparse histogram mechanism behaves identically to the Gaussian mechanism with a post-processing threshold $\tauhi$ applied to the count column.  Utilizing Observation~\ref{lemma:no_positive_requirement} to account for possibly negative $\epsilon_3$, we can then claim where the right-hand side is the evaluation for the Gaussian mechanism without post-processing: 
\begin{lemma}
We have that
\begin{align}
&\Prob(L^{+}_{+} \geq \epsilon_2) - e^{\epsilon_2} \Prob(L^{+}_{-} \leq -\epsilon_2) \leq \Phi\left( \frac{\mu}{2} - \frac{\epsilon_2}{\mu}\right) - e^{\epsilon_2} \Phi\left(-\frac{\mu}{2} - \frac{\epsilon_2}{\mu}\right)\nonumber\\
&\Prob(L^{+}_{-} \geq \epsilon_3) - e^{\epsilon_3} \Prob(L^{+}_{+} \leq -\epsilon_3) \leq \Phi\left(\frac{\mu}{2} - \frac{\epsilon_3}{\mu}\right) - e^{\epsilon_3} \Phi\left(-\frac{\mu}{2} - \frac{\epsilon_3}{\mu}\right)
\label{eq:plrv_simplification}
\end{align}
where the functions $\epsilon_2$ and $\epsilon_3$ are defined in Lemma~\ref{lemma:aplus} and this $\mu$ is
\begin{align}
&\mu(a_{+}) = \sqrt{\frac{a_{+}}{\sigma^2} + a_{+}\mu_o^2}.
\label{eq:mu_aplus}
\end{align}
Without further assumptions about $A$ and the groups, these inequalities are tight.
\label{lemma:unbounded_consequence}
\end{lemma}

Combining the above lemmas, and that the quantity in Equation~\ref{eq:analytic_gaussian} is a monotonically increasing function of $\mu$, gives our final result

\begin{theorem}
Algorithm~\ref{algo:gshm} with parameters $\tauhi$, $\taulow$, $\sigma$, and $\Sigma$ satisfies ($\epsilon$, $\delta$)-DP if the following condition holds
\begin{align}
&\max \biggr[ 1 - \Phi\left(\frac{\tau^{*} - \tau}{\sigma}\right)^{C_u}, \nonumber\\
&\max_{a_{+} + a_{=} = C_u, a_{+} > 0} 1 - \Phi\left(\frac{\tau^{*} - \tau}{\sigma}\right)^{a_{=}} + \Phi\left(\frac{\tau^{*} - \tau}{\sigma}\right)^{a_{=}}\left[\Phi\left( \frac{\mu}{2} - \frac{\epsilon_2}{\mu}\right) - e^{\epsilon_2} \Phi\left(-\frac{\mu}{2} - \frac{\epsilon_2}{\mu}\right)\right], \nonumber\\
&\max_{a_{+} + a_{=} = C_u, a_{+} > 0} \Phi\left(\frac{\mu}{2} - \frac{\epsilon_3}{\mu}\right) - e^{\epsilon_3} \Phi\left(-\frac{\mu}{2} - \frac{\epsilon_3}{\mu}\right) \biggr] \leq \delta
\label{eq:final_result}
\end{align}
where the functions $\epsilon_2$ and $\epsilon_3$ are defined in Lemma~\ref{lemma:aplus}, and the function $\mu$ is defined in Lemma~\ref{lemma:unbounded_consequence}.  

Without further assumptions on A and the groups, this privacy accounting is exact. 
\label{th:gshm_privacy}
\end{theorem}

Because the optimization is over $\aplus + \aequals = \usermax$, this expression can be evaluated in linear-time with respect to $\usermax$.  Now let us compare our result in Theorem~\ref{th:gshm_privacy} against ``add the deltas'' Theorem~\ref{th:add_the_deltas} directly.  
\begin{corollary}
Let $\mu(\usermax)$ be Eq.~\ref{eq:mu_aplus} evaluated at $\usermax$ and define $m \geq 1$ generalizations of Eq.~\ref{eq:m1_deltas}:
\begin{align}
\delta_{\text{Gaussian}} &= \Phi\left(\frac{\mu(\usermax)}{2} - \frac{\epsilon}{\mu(\usermax)}\right) - e^{\epsilon} \Phi\left(-\frac{\mu(\usermax)}{2} - \frac{\epsilon}{\mu(\usermax)}\right) \nonumber\\
\delta_{\text{infinite}} &= 1 - \Phi\left(\frac{\tauhi-\taulow}{\sigma}\right)^{\usermax}.
\end{align}
Algorithm~\ref{algo:gshm} with parameters $\tauhi$, $\taulow$, $\sigma$, and $\Sigma$ has a minimal $\delta$ at a given $\epsilon \geq 0$, given by equality in Eq.~\ref{eq:final_result}, where
\begin{align} 
\max(\delta_{\text{infinite}}, \delta_{\text{Gaussian}}) \leq \delta < \delta_{\text{infinite}} + \delta_{\text{Gaussian}}.
\label{eq:upper_and_lower_bound_on_delta}
\end{align}
\label{th:gshm_corollary}
For $\usermax=1$, the lower-bound is an equality.
\end{corollary}
With realistic parameters, the minimal $\delta$ is often equal to the lower-bound in this corollary.  Equality with the lower-bound both implies no additional privacy cost for thresholding over the Gaussian mechanism with the same noise and $\epsilon$ when $\delta_{\text{infinite}} \leq \delta_{\text{Gaussian}}$, and a separation from the upper-bound of $\delta_{\text{infinite}} + \delta_{\text{Gaussian}}$, the bound for $m=1$ in Theorem~\ref{th:add_the_deltas} derived by Wilson et al. \cite{wilson_et_al}.

Because our analysis simplifies at $\usermax=1$, we can derive a precise comparison between the minimum noisy threshold $\tauhi-\taulow$ required between using ``add the deltas" versus our improved accounting.  Recall that we want to use the smallest $\tauhi - \taulow$ to preserve utility.
\begin{corollary}
Let $\usermax = 1$ and $\delta \geq \delta_{\text{Gaussian}}$.  Then the ratio of the minimal $\tauhi - \taulow$ difference that satisfies $(\epsilon, \delta)$-DP for Algorithm~\ref{algo:gshm} with other parameters $\sigma$ and $\Sigma$ under ``add the deltas'' and exact accounting is given by
\begin{align}
\frac{\Phi^{-1}(1-\delta+ \delta_{\text{Gaussian}})}{\Phi^{-1}(1-\delta)}
\end{align}
\label{cor:ratio_threshold}
\end{corollary}
This ratio is always greater than one and can be arbitrarily large, implying arbitrarily large gains in utility due to smaller noisy thresholds are possible via the exact accounting for fixed privacy parameters.  We shall see similar behavior when $\usermax > 1$ in our case study.

\section{Case study on URL dataset}\label{sec:alt_case_study}
Differential privacy implementations for count datasets with grouping columns typically require constructing a Cartesian expansion across all combinations of values in the grouping columns that are not \emph{structural zeros} (i.e., impossible combinations, like 98-year-old infants). This can require the inclusion of a very large number of rows in a private dataset that are ``sampling zeroes'' (counts that happen to be zero in the dataset but are not structural zeros) which become indistinguishable from small positive values after the addition of noise.  Let us consider an example implementation of the Gaussian sparse histogram method using the exact accounting in this paper, as compared to the ``add the deltas'' accounting.

We consider differences using the ``Facebook Privacy-Protected Full URLs Data Set,'' which we will refer to as the Facebook URL Shares dataset (for more details on this dataset, see \citep{facebook_urls}). The Facebook URL Shares dataset contains aggregated and de-identified information about exposure to and engagement with URLs that were shared on Facebook. The key table of data in this dataset is called the ``URL Breakdowns'' table, which has columns recording the number of users who viewed, clicked, liked, reacted, commented, or shared any URL that had been posted to Facebook, provided that URL had been shared publicly at least 100 times.\footnote{Note that Laplace noise was added to the public share counts for each URL prior to implementing the 100 public shares threshold, so this was only post-processing.} Gaussian noise was added to each of the count columns in order to satisfy action-level and user-level differential privacy (the former protecting user interactions with a particular URL in the dataset and the latter protecting a user's cumulative interactions with URLs in the dataset). The differential privacy implementation was set such that the 99th percent most active user would receive a specified $(\epsilon, \delta)$ privacy guarantee. 

The URL Breakdowns table groups URL engagement data columns based on: (1) year and month when the interaction took place (2) six user age brackets plus a \verb+NULL+ category (3) user gender (4) user country of residence and (5) a 5 category user ``political page affinity'' categorization, plus a \verb+NULL+ category, for U.S. users only. The privacy implementation for the URL Breakdowns table was not via a sparse histogram method, and required constructing a Cartesian expansion across all five aggregation columns. That meant that in the initial dataset covering 31 year-months and 46 countries, every URL included in the dataset would have 33,201 rows in the breakdowns table: 29,295 rows for all non-U.S. countries (45 countries, 31 year-months, 7 age categories, and 3 gender categories) and 3,906 for the U.S. These rows need to be included for each URL in the dataset, even if a given URL only received engagement in one country across one year month.

A sparse histogram mechanism would allow us to exclude all rows with true values of zero, but at the cost of setting a noisy threshold that would filter out some non-zero values. The URL views column would present a logical choice as a filter column, because other types of interactions can only occur if a URL has been viewed (i.e. users can't click or like a URL they have not seen). In theory, a URL in this dataset could have zero views, but this would be highly unlikely for any URL that received over 100 public shares. As discussed in the dataset codebook, each interaction column limits users to contributing one interaction per column per row, so the data is already structured in a manner that would make it well suited to implement our mechanism.

The codebook notes that the 99th percent most active user contributed $51,914$ URL views (a procedure was used to compute a noisy version of this statistic, see \cite{facebook_urls} for more details).  Following the privacy guarantee aimed at the 99th percent most active user, we set $\usermax=51,914$ for our case study.  Because ``add the deltas" was previously derived for the $m=1$ setting, we limit our case study to just considering the views column.  Incorporating the other columns (via a non-zero $\mu_o^2$) only increases differences between accounting methods.  The codebook notes that the standard deviation of the Gaussian noise added to the views column was $\sigma=2228$, which as a Gaussian mechanism satisfies $(\epsilon = 0.349, \delta=10^{-5})$-DP according to Theorem~\ref{th:analytic_gaussian}.  

Let us say that we are interested in the Gaussian sparse histogram mechanism for implementing differential privacy for the views column in the Facebook URL Shares dataset, with $\taulow=1$.  After fixing $\taulow$, the Gaussian sparse histogram mechanism has two remaining parameters $\tauhi$, $\sigma$.  We will consider two scenarios; the minimal $\tauhi$ versus $\sigma$ that satisfies a given  $(\epsilon, \delta)$-DP constraint, and $(\epsilon, \delta)$-DP curves for a fixed $\sigma$ and $\tauhi$.  In both cases, we will see a separation between the curves produced by the exact and ``add the deltas'' accounting.

\begin{figure}[t]
    \centering
    \begin{subfigure}{0.49\textwidth}
        \centering
        \includegraphics[width=0.95\textwidth]{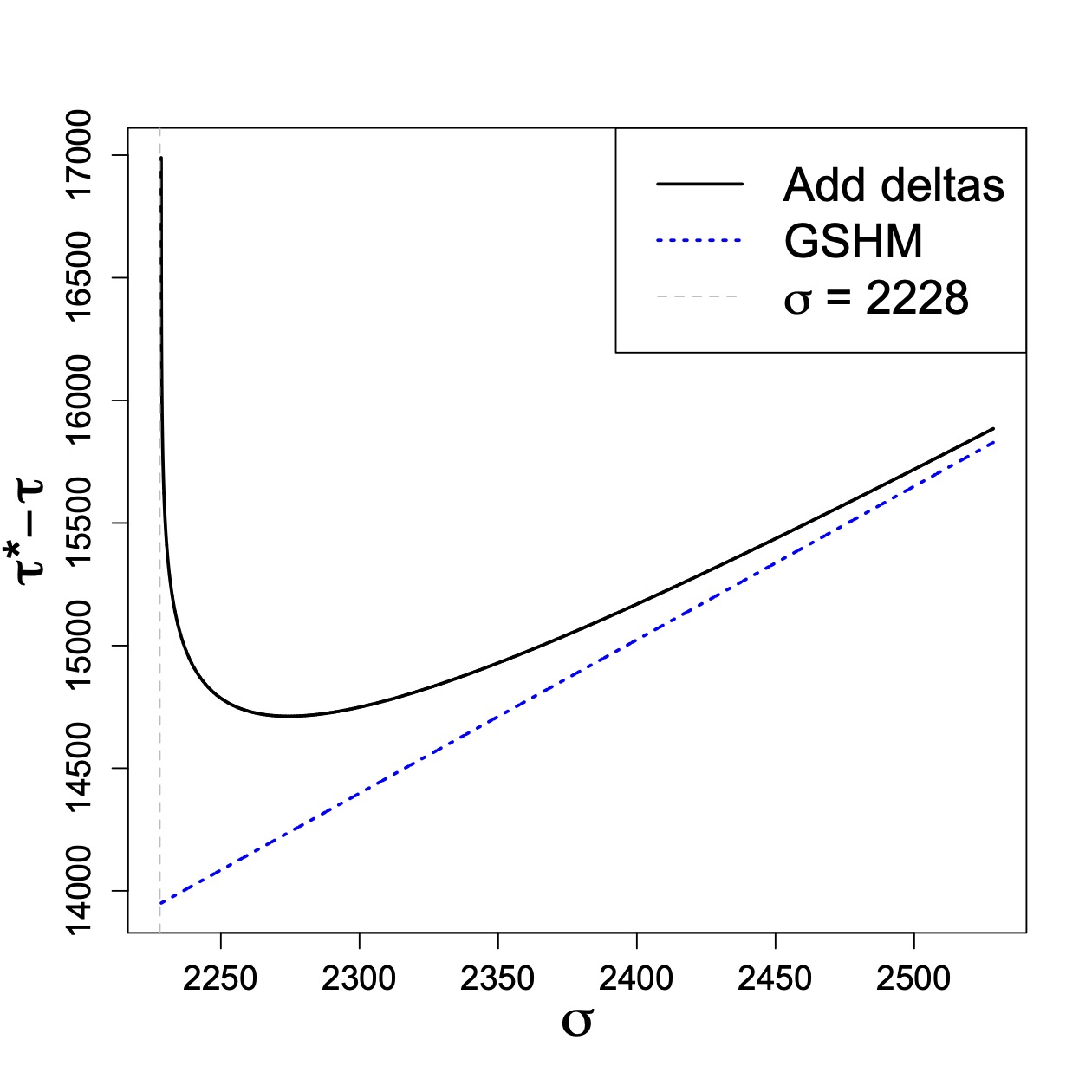}
        \caption{\label{fig:sigmatau}}
    \end{subfigure}
    \hfill
    \begin{subfigure}{0.49\textwidth}
        \centering
        \includegraphics[width=0.95\textwidth]{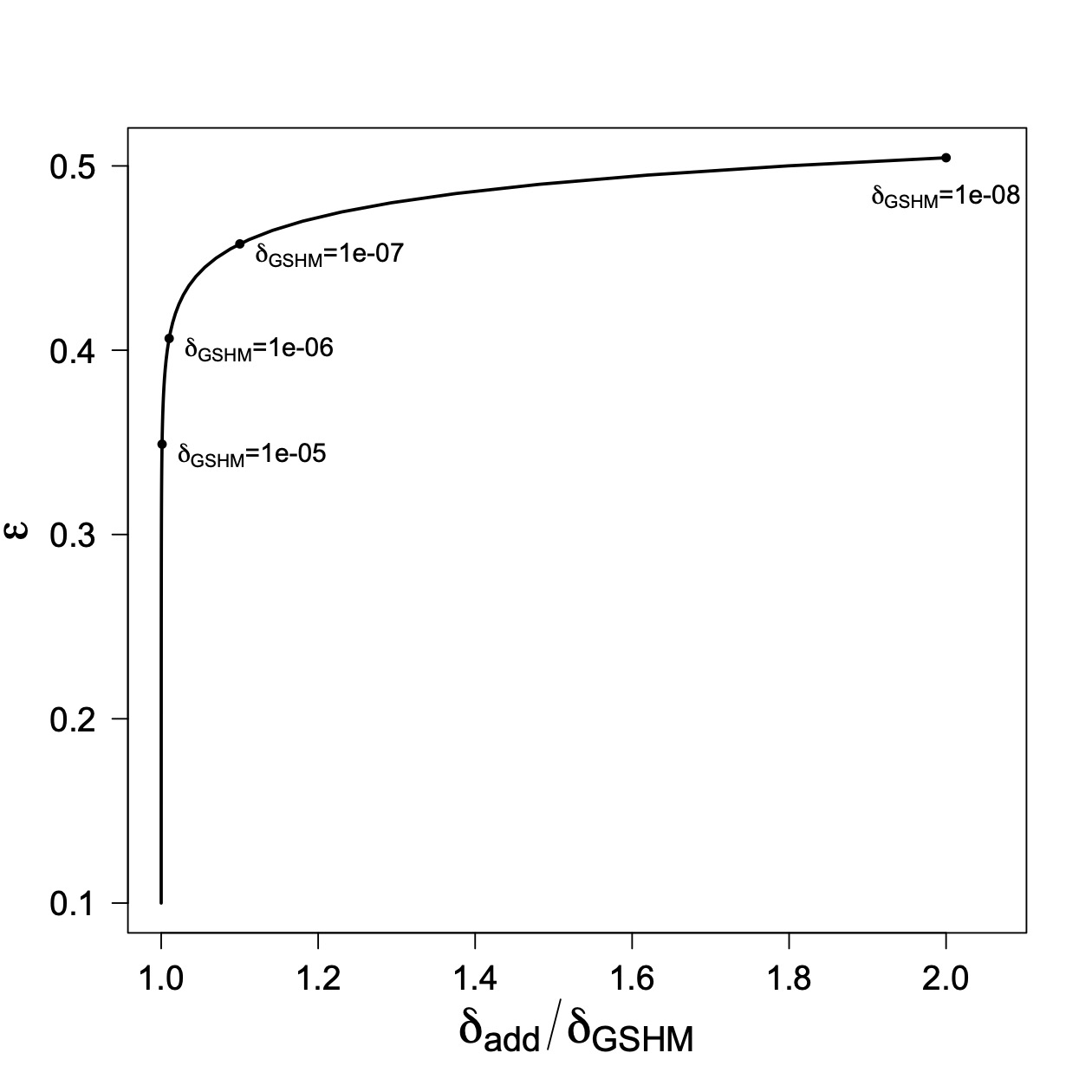}
        \caption{\label{fig:epsdelta_curve}}
    \end{subfigure}
    \caption{(a). Scenario 1 showing the minimal $\tauhi$ versus $\sigma$ satisfying $(\epsilon = 0.349, \delta = 10^{-5})$-DP for both ``add the deltas'' and exact accounting. (b). Scenario 2 showing a $(\delta_{\text{add}}/\delta_{\text{GSHM}}, \epsilon)$ curve for $\sigma=2228$ and $\tauhi - \taulow = 16176$, where $\delta_{\text{add}}$ is from ``add the deltas'' and $\delta_{\text{GSHM}}$ is the exact accounting.  The ``add the deltas'' and exact accounting is in Theorem~\ref{th:add_the_deltas} and Theorem~\ref{th:gshm_privacy} respectively.}
\end{figure}

\subsection{Scenario 1:  Comparison of minimal $\tauhi$ versus $\sigma$}

We fix $(\epsilon = 0.349, \delta=10^{-5})$, the same privacy parameters implied by the Gaussian mechanism for the views column.  As the Gaussian sparse histogram mechanism cannot release a lower $\sigma$ than the Gaussian mechanism at the same privacy, we therefore consider $\sigma \geq 2228$.  At $\sigma=2228$, the Gaussian sparse histogram mechanism can use $\tau^* = 13948$.~\footnote{Whether the Gaussian sparse histogram mechanism improves upon simply post-processing the existing Facebook URL Shares dataset released via the Gaussian mechanism is unlikely in this circumstance.  Consider dropping rows with noisy view counts less than some desired sparsity threshold $\tau_{post}$. If $\tau_{post}=\tauhi=13948$, the probability of a given zero row remaining after post-processing is roughly $10^{-10}$.  This probability is extremely small indicating a smaller $\tau_{post}$ would likely suffice for sparsity.  On the other hand, if a desired $\tau_{post} \geq \tauhi$, the Gaussian sparse histogram mechanism is preferred as it performs the same filtering on non-zero count rows, while removing the zero count rows.  Our emphasis in this case study is to understand the effects of privacy accounting, not to determine whether applying the Gaussian sparse histogram method would have produced a more useful dataset.}  For each $\sigma$, we compute the minimum $\tauhi$ that satisfy $(\epsilon = 0.349, \delta=10^{-5})$-DP from ``add the deltas`` in Theorem~\ref{th:add_the_deltas} and our exact accounting in Theorem~\ref{th:gshm_privacy}.  We show the resulting curves in Figure~\ref{fig:sigmatau}.

We see that the exact accounting curve produces strictly lower thresholds $\tauhi$ than ``add the deltas''.  The difference is greatest as we approach $\sigma=2228$ where ``add the deltas'' cannot produce a threshold $\tauhi$ that meets the criteria at this lower-bound.  Like in Corollary~\ref{cor:ratio_threshold}, the difference at $\sigma=2228$ is unbounded, and it is precisely these lowest $\sigma$ and lowest $\tauhi$ values that are of primary interest as they provide the maximum utility.  The shape of the exact accounting curve requires no tradeoff between the two objectives, as we can choose both the lowest $\sigma$ and the lowest $\tauhi$.  However, for ``add the deltas'' we are required to use a higher $\sigma$ and higher $\tauhi$ to satisfy the criteria.  We can further quantify these differences in noisy threshold in terms of the number of additional non-zero rows removed due to using a higher threshold\footnote{For these calculations, we use the breakdowns table in the Facebook URL Shares dataset that covers a period from January 2017 to February 2021, for users living in the U.S. The URL Shares dataset is updated periodically as new data become available. An exact computation of the expected fraction of rows lost as a function of $\sigma$ and $\tauhi$ would require access to data not included in the Facebook URL Shares dataset. But in practice, the expected fraction of rows lost should be almost identical when computed using the privacy-protected version of the Facebook URL Shares dataset that is available to researchers via Social Science One. This is because after dropping rows of the privacy-protected data where noisy views are smaller than $\tauhi$, the number of zero-valued rows is vanishingly small. The privacy-protected dataset has on the order of $2 \times 10^{11}$ rows, of which over 38 million rows have more than $14022$ noisy views (where country is U.S. and where the views occurred between January 2017 and February 2021). If all the rows had true values of zero (with $2 \times 10^{11}$ rows, $\sigma = 2240$ and $\tauhi = 14022$), the expected number of zero-valued rows with noisy views greater than $\tauhi$ is $31$, which is vanishingly small considering the over 38 million rows with noisy views greater than $14022$.}

\begin{itemize}
    \item With $\sigma=2400$, we will lose about 1.1\% more non-zero rows
    \item With $\sigma=2300$, we will lose about 2.8\% more non-zero rows
    \item With $\sigma=2240$, we will lose about 6.9\% more non-zero rows
\end{itemize}

The two curves converge as $\sigma \to \infty$. ``Add the deltas'' gets within 1\% of Theorem ~\ref{th:gshm_privacy} at $\sigma=2396$ (the necessary value of $\tau^*-\tau$ is $15,148$ under ``add the deltas'' and $14,998$ under Theorem ~\ref{th:gshm_privacy}). ``Add the deltas'' gets within 0.1\% of Theorem ~\ref{th:gshm_privacy} at $\sigma = 2699$ (the necessary value of $\tau^*-\tau$ is about $16,910$ under ``add the deltas'' and $16,894$ Theorem ~\ref{th:gshm_privacy}).  At the points where the curves converge though, we are adding much more noise and thresholding out far more rows than would be necessary to satisfy the desired differential privacy target.  Further, this convergence occurs only because we examined just the views column (i.e. $m=1$ and $\mu_o = 0$).  If additional columns were included, the curves may converge to a constant factor as in Corollary~\ref{cor:ratio_threshold}, considered as $\sigma$ varies.

\subsection{Scenario 2:  Comparison of ($\epsilon$, $\delta$) curve}

We can also fix $\sigma$ and $\tauhi$ and examine the ($\epsilon$, $\delta$)-DP curves produced by ``add the deltas'' and our exact accounting.  For this example, say we use $\sigma=2228$ and $\tauhi-\taulow = 16176$ which meets $\delta_{\text{infinite}} = 10^{-8}$.  For our curves, we know that $\delta \geq \delta_{\text{infinite}}$.  

With these parameters set, we can examine $\delta(\epsilon)$ or $\epsilon(\delta)$.  We consider the former in Figure~\ref{fig:epsdelta_curve} which displays how $\delta(\epsilon)$ varies over $\epsilon \in [0.1,0.504]$ under Theorem~\ref{th:gshm_privacy} (GSHM) and ``add the deltas'' under Theorem~\ref{th:add_the_deltas}.  As seen in Figure~\ref{fig:epsdelta_curve}, the final $\delta$ returned by ``add the deltas'' is double that of GSHM at $\delta_{\text{GSHM}} = 10^{-8}$, 10\% greater at $\delta_{\text{GSHM}} = 10^{-7}$, 1\% greater at $\delta_{\text{GSHM}} = 10^{-6}$, and 0.1\% greater at $\delta_{\text{GSHM}} = 10^{-5}$.  This aligns with our expectations from Corollary~\ref{th:gshm_corollary}.  When $\delta_{\text{infinite}}$ is very small with respect to $\delta_{\text{Gaussian}}$, the lower and upper-bounds in Eq.~\ref{eq:upper_and_lower_bound_on_delta} become closer and $\delta$ produced from both accounting approaches will become similar.  However when $\delta_{\text{infinite}}$ is non-trivial compared to $\delta_{\text{Gaussian}}$, the exact accounting produces a smaller $\delta$, by up to a factor of two.

\section{Conclusion}\label{sec:conclusion}
Applications of differential privacy to count datasets traditionally require constructing a Cartesian expansion across all possible combinations of values in grouping columns.  Constructing such a Cartesian expansion can be difficult or impossible for multiple reasons, especially when the domains are large or even infinite.  In these cases, sparse histogram methods provide reasonable alternatives to Cartesian expansion.

In this paper, we have provided an exact privacy loss analysis of the Gaussian sparse histogram mechanism and demonstrated that our exact accounting was feasible in practice.  On our URL case study, our comparison against past research demonstrated that in practical circumstances our more precise privacy accounting can increase utility by a significant amount, primarily in situations where it is desirable to set a low enough noisy threshold such that $\delta_{\text{infinite}}$ is comparable to $\delta_{\text{Gaussian}}$.  On the other hand, when $\delta_{\text{infinite}}$ can be made vanishing through use of a large noisy threshold, our accounting matches those from ``add the deltas''.  Given that the implementation of exact accounting is simple and that smaller noisy thresholds are of primary concern when using a sparse histogram method, we believe our improved accounting should be useful in practice.

The exactness of our privacy analysis relies upon uniform sensitivity across groups for $m > 1$ and unbounded group counts. If the count of users per group is bounded or non-uniformity is of interest, future research could improve upon our privacy analysis via revisiting Lemma~\ref{lemma:unbounded_consequence} with additional assumptions.

\textit{Acknowledgements:} This research was supported by funding from Meta.

\bibliographystyle{plain}
\bibliography{thresholding_works_cited}

\section{Appendix}\label{sec:appendix}
\subsection{Additional preliminaries for privacy analysis}
We partition the output space $\Omega$ into events for which $L_{+}$ (through $L^{=}_{+}$) is finite and infinite.  Let $F_j$ be all events where some subset of the rows $a_{=}$ are not $\Null^{\numcols}$. $M(X)\in F_j$ if and only if $L_{+}$ is infinite. 
We have these facts:
\begin{itemize}
    \item $\Prob_{M(X_{-j})}( F_j)=0$
    \item $\Prob_{M(X_{-j})}(\sim F_j)=1$
    \item $\Prob_{M(X)}( F_j)=1-\Phi\left(\frac{\tau^*-\tau}{\sigma}\right)^{a_=}$
        \item $\Prob_{M(X)}(\sim F_j)=\Phi\left(\frac{\tau^*-\tau}{\sigma}\right)^{a_=}$
    \item $L^{=}_+ = \infty$ with probability $1-\Phi\left(\frac{\tau^*-\tau}{\sigma}\right)^{a_=}$ and equals $\log\frac{\Phi\left(\frac{\tau^*-\tau}{\sigma}\right)^{a_=}}{1}=a_=\log(\Phi\left(\frac{\tau^*-\tau}{\sigma}\right))$ with probability $\Phi\left(\frac{\tau^*-\tau}{\sigma}\right)^{a_=}$
    \item $L^{=}_- = \log\frac{1}{\Phi\left(\frac{\tau^*-\tau}{\sigma}\right)^{a_=}}=-a_=\log(\Phi\left(\frac{\tau^*-\tau}{\sigma}\right))>0$ with probability 1.
\end{itemize}

\subsection{Lemma~\ref{lemma:aequals}:  PLRV evaluation when $a_+=0$}
In this case, we have the following facts:
\begin{itemize}
    \item $L^{+}_{+}=0$
    \item $L^{+}_{-}=0$
    \item For any $\epsilon>0$, $\Prob(L^=_+ \geq \epsilon)=1-\Phi\left(\frac{\tau^*-\tau}{\sigma}\right)^{a_=}$ and $\Prob(L^=_-\leq -\epsilon)=0$
    \item For any $\epsilon>0$, $\Prob(L^=_- \geq \epsilon)=1$ if $\epsilon\leq -a_=\log(\Phi\left(\frac{\tau^*-\tau}{\sigma}\right))$ and 0 otherwise. $\Prob(L^=_+ \leq -\epsilon)=\Phi\left(\frac{\tau^*-\tau}{\sigma}\right)^{a_=}$ if $-\epsilon \geq a_=\log(\Phi\left(\frac{\tau^*-\tau}{\sigma}\right))$ and is $0$ otherwise.
\end{itemize}

So
\begin{align*}
    \Prob(L^=_+\geq \epsilon) - e^{\epsilon} \Prob(L^=_- \leq -\epsilon) &= 1-\Phi\left(\frac{\tau^*-\tau}{\sigma}\right)^{a_=}\\
    \Prob(L^=_-\geq \epsilon) - e^{\epsilon} \Prob(L^=_+ \leq -\epsilon) &=\begin{cases}
    0 & \text{ if $\epsilon > -a_=\log(\Phi\left(\frac{\tau^*-\tau}{\sigma}\right))$}\\
    1-e^{\epsilon}\Phi\left(\frac{\tau^*-\tau}{\sigma}\right)^{a_=} & \text{ if } \epsilon \leq -a_=\log(\Phi\left(\frac{\tau^*-\tau}{\sigma}\right))
    \end{cases}
\end{align*}
Note that the max of these is $1-\Phi\left(\frac{\tau^*-\tau}{\sigma}\right)^{a_=}$ and maximizing over $a_=\leq C_u$ we get 
\begin{align}
1-\Phi\left(\frac{\tau^*-\tau}{\sigma}\right)^{C_u} \leq \delta
\end{align}

\subsection{Lemma~\ref{lemma:aplus}:  PLRV evaluation when $\aplus >0$}

Conditioned on event $F_j$ not happening, we have
\begin{align}
&L_{+} | \{M(X)\in\sim F_j\} = L^{+}_{+} + \log \frac{\Prob_{M(X)}(\sim F_j)}{\Prob_{M(X_{-j})}(\sim F_j)} = L^{+}_{+} + a_{=} \log \Phi\left(\frac{\tau^{*} - \tau}{\sigma}\right)  \nonumber\\
&L_{-} | \{M(X_{-j})\in\sim F_j\} = L^{+}_{-} +\log\frac{\Prob_{M(X_{-j})}(\sim F_j)}{ \Prob_{M(X)}(\sim F_j)} = L^{+}_{-} - a_{=} \log\Phi\left(\frac{\tau^{*} - \tau}{\sigma}\right).
\end{align}

As mentioned in the main text, for our mechanism to be $(\epsilon, \delta)$-DP, we require the following two expressions hold for any values of $a_{+} + a_{=} \leq C_u$.
\begin{align}
\Prob(L_{+} \geq \epsilon) - e^{\epsilon} \Prob(L_{-} \leq -\epsilon) \leq \delta \nonumber\\
\Prob(L_{-} \geq \epsilon) - e^{\epsilon} \Prob(L_{+} \leq -\epsilon) \leq \delta
\end{align}
The first expression corresponds to $X$ having $j$ and $X'$ not having $j$, 
and the second expression corresponds to $X$ not having $j$ and $X'$ having $j$. 

Let's first consider the top expression.  Dividing into conditioning on $\sim F_j$ and $F_j$, and remembering that $\sim F_j$ happens with probability one under $M(X_{-j})$, we work this out to be
\begin{align}
&\Prob_{M(X)}(F_j) \Prob(L_{+} \geq \epsilon | F_j) + \Prob_{M(X)}(\sim F_j) \Prob(L_{+} \geq \epsilon | \sim F_j) - e^{\epsilon} \Prob(L_{-} \leq -\epsilon | \sim F_j) = \nonumber\\
&\Prob_{M(X)}(F_j) + \Prob_{M(X)}(\sim F_j) \Prob(L_{+} \geq \epsilon | \sim F_j) - e^{\epsilon} \Prob(L_{-} \leq -\epsilon | \sim F_j) = \nonumber\\
&1 - \Prob_{M(X)}(\sim F_j) + \Prob_{M(X)}(\sim F_j) \left[\Prob(L_{+} \geq \epsilon | \sim F_j) - e^{\epsilon - \log(\Prob_{M(X)}(\sim F_j))} \Prob(L_{-} \leq -\epsilon | \sim F_j)\right] = \nonumber\\
&\underbrace{1 - \Phi\left(\frac{\tau^{*} - \tau}{\sigma}\right)^{a_{=}}}_{\text{$\delta$ for infinite privacy loss}} + \underbrace{\Phi\left(\frac{\tau^{*} - \tau}{\sigma}\right)^{a_{=}} [ \Prob(L^{+}_{+} \geq \epsilon_2) - e^{\epsilon_2} \Prob(L^{+}_{-} \leq -\epsilon_2)]}_{\text{$\delta$ from the numerical $a_+$ rows, no $a_=$ rows are output}} \nonumber
\end{align}
where $\epsilon_2 = \epsilon - a_{=} \log \Phi\left(\frac{\tau^{*} - \tau}{\sigma}\right)$. Note that $\epsilon>0\Rightarrow\epsilon_2>0$.

Now let's return to the bottom expression that considers $X$ without $j$ and $X'$ with $j$.  For this case, we want
\begin{align}
\Prob(L_{-} \geq \epsilon) - e^{\epsilon} \Prob(L_{+} \leq -\epsilon) \leq \delta.
\end{align}
Expanding out the left-hand side (similar to before, being careful with the signs and noting that (a) $\Prob(L_{+} \leq - \epsilon | F_j) = 0$ and (b) $\Prob(L_{-} \geq \epsilon | \sim F_j)=\Prob(L_{-} \geq \epsilon )$ (since $\sim F_j$ always happens under $M(X_{-j}))$, we get
\begin{align}
&\Prob(L_{-} \geq \epsilon | \sim F_j) - e^{\epsilon} \biggr(\Prob_{M(X)}(\sim F_j) \Prob(L_{+} \leq -\epsilon | \sim F_j) + \Prob_{M(X)}(F_j) \Prob(L_{+} \leq -\epsilon | F_j)\biggr) \nonumber\\
&= \Prob(L_{-} \geq \epsilon | \sim F_j) - e^{\epsilon + \log(\Prob(\sim F_j))}  \Prob(L_{+} \leq -\epsilon | \sim F_j) \nonumber\\
&=\Prob(L^{+}_{-} \geq \epsilon_3) - e^{\epsilon_3} \Prob(L^{+}_{+} \leq -\epsilon_3)
\end{align}
where $\epsilon_3 = \epsilon + a_{=} \log \Phi\left(\frac{\tau^{*} - \tau}{\sigma}\right)$.

\subsection{Lemma~\ref{lemma:unbounded_consequence}}

Let $A_{+}$ be the set of $\aplus$ rows containing $j$ with counts greater than the threshold $\taulow$.  To evaluate the remaining PLRV expressions when $\aplus > 0$, for these $A_{+}$ rows, we note that the Gaussian sparse histogram mechanism applied to these rows is identical to the Gaussian mechanism with a post-processing threshold $\tauhi$ applied to the noisy counts for each row.  Lemma~\ref{lemma:unbounded_consequence} then follows from the following argument about post-processing which says we can use the PLRV expressions from the Gaussian mechanism as an upper-bound, regardless of the sign of $\epsilon$.  Observation~\ref{lemma:no_positive_requirement} allows utilizing the Gaussian PLRV expressions contained in Theorem~\ref{th:analytic_gaussian}, despite possibly negative $\epsilon_3$.   Finally, since we are releasing these rows each of which has a $\mu_i$ contribution, the total $\mu^2$ for releasing the $A_{+}$ rows is given by $\sum_{i \in A_{+}} \mu_i^2 \leq \frac{\aplus}{\sigma^2} + \aplus \mu_o^2$.  Recalling that the Gaussian PLRV expressions are increasing functions of $\mu$, using a larger $\mu^2$ (and therefore larger $\mu$) results in a larger upper-bound.  

Under uniformity ($\mu_i^2$ contributions equal for all rows $i$), this final inequality for the $\mu$ contribution from the $A_{+}$ rows is an equality.  Without any assumptions on $A(x)$ and the groups, uniformity is possible (and reasonable in many circumstances) and hence this inequality is tight.  Let us now also consider the tightness of the inequality due to post-processing.  Consider a pair of neighboring datasets $\X$ and $\Xminusj$ where for all rows in $A_{+}$ the counts in both $\X$ and $\Xminusj$ are very large compared to the threshold $\tauhi$, such that the chance of a noisy user count being less than $\tauhi$ goes to zero.  Therefore the privacy loss random variables over the $A_{+}$ rows can behave arbitrarily close to the Gaussian mechanism by simply considering datasets with large enough counts on these rows.  Hence there exists a pair of neighboring datasets $\X$ and $\Xminusj$ such that the PLRV expressions from the Gaussian mechanism are arbitrarily close to those of applying the Gaussian sparse histogram mechanism on these $A_{+}$ rows.

\subsubsection{Post-processing and PLRV's}

We modify \cite{balle_wang}'s proof of theorem 5 to prove the following claim.   

Let $M$ be a random function from $O$ to $R$. Let $f$ be a deterministic post-processing function from $R$ to $R'$.  Then for any datasets $\X$ and $\X'$, and any value of $\epsilon$ including $\epsilon < 0$, we have that
\begin{align}
\Prob(L_{f \circ M, \X, \X'} \geq \epsilon) - e^{\epsilon} \Prob(L_{f \circ M, \X', \X} \leq -\epsilon) \leq \Prob(L_{M, \X, \X'} \geq \epsilon) - e^{\epsilon} \Prob(L_{M, \X', \X} \leq -\epsilon).
\end{align}
\textbf{Proof:}  Let $T = \{r' \in R': \log[\frac{\Prob(f(M(\X)) = r')}{\Prob(f(M(\X')) = r')}] \geq \epsilon\}$, and let $S = \{r \in R: f(r) \in T\}$.  Let $E = \{r \in R: \log[\frac{\Prob(M(\X) = r)}{\Prob(M(\X') = r)}] \geq \epsilon\}$.  Also $E_{+} = S \cap E$ and $E_{-} = S \cap (R / E)$.   Using these definitions we can write
\begin{align}
\Prob(L_{f \circ M, \X, \X'} \geq \epsilon) - e^{\epsilon} \Prob(L_{f \circ M, \X', \X} \leq -\epsilon) &= \int_{T} \Prob(f(M(\X)) = r') - e^{\epsilon} \Prob(f(M(\X')) = r') dr' \nonumber\\
&= \int_{S} \Prob(M(\X) = s) - e^{\epsilon} \Prob(M(\X') = s) ds \nonumber\\
&= (\int_{E_{+}} + \int_{E_{-}}) \Prob(M(\X) = s) - e^{\epsilon} \Prob(M(\X') = s) ds \nonumber\\
&\leq \int_{E_{+}} \Prob(M(\X) = s) - e^{\epsilon} \Prob(M(\X') = s) ds \nonumber\\
&\leq \int_{E} \Prob(M(\X) = s) - e^{\epsilon} \Prob(M(\X') = s) ds \nonumber\\
&= \Prob(L_{M, \X, \X'} \geq \epsilon) - e^{\epsilon} \Prob(L_{M, \X', \X} \leq -\epsilon),
\end{align}
where we used that under the events in $E_{-}$ the contributions are all non-positive for the first inequality, and then the contributions under any events in $E$ are non-negative and $E_{+} \subseteq E$.

\subsection{Theorem~\ref{th:gshm_privacy}}

The three terms in the Theorem immediately follow from Lemma~\ref{lemma:aequals}, Lemma~\ref{lemma:no_positive_requirement}, and Lemma~\ref{lemma:unbounded_consequence}.  The only remaining aspect is to prove that the inner maximization occurs when $\aplus + \aequals = \usermax$ instead of $\aplus + \aequals \leq \usermax$.  To do so, we demonstrate that the PLRV difference is monotone with respect to $\mu$ for any $\epsilon$.  This implies equality with $\usermax$ because $\aplus$ only enters into these expressions via $\mu(\aplus)$ and $\mu(\aplus)$ in Equation~\ref{eq:mu_aplus} is monotonically increasing in $\aplus$.
\begin{corollary}
\label{corollary:bw_monotone}
The left-hand side of Eq.~\ref{eq:analytic_gaussian} is monotonically increasing with respect to $\mu$. 
\end{corollary}

\noindent \textbf{Proof of Corollary~\ref{corollary:bw_monotone}} \\
Let
\begin{align} 
f(\mu, \epsilon) = \Phi\left( \frac{\mu}{2} - \frac{\epsilon}{\mu}\right) - e^{\epsilon} \Phi\left(-\frac{\mu}{2} - \frac{\epsilon}{\mu}\right).
\label{eq:f}
\end{align}
for arbitrary $\mu > 0$ and $\epsilon$.  Applying calculus, we have that
\begin{align}  
\frac{\partial f}{\partial \mu} &= \phi\left(\frac{\mu}{2} - \frac{\epsilon}{\mu}\right), \\
\frac{\partial f}{\partial \epsilon} &= -e^{\epsilon} \Phi\left(-\frac{\mu}{2} - \frac{\epsilon}{\mu}\right)
\end{align}
where $\phi$ is the PDF of the standard normal distribution.  So the partial derivative of $f$ with respect to $\mu$ is always positive, and the partial derivative with respect to $\epsilon$ is always negative.  

\subsection{Corollary~\ref{th:gshm_corollary}}

For concreteness, we recall the $m \geq 1$ versions of $\delta_{\text{Gaussian}}$ and $\delta_{\text{infinite}}$, where $\mu(\usermax)$ is the function defined in Eq.~\ref{eq:mu_aplus} evaluated at $\usermax$
\begin{align}
\delta_{\text{Gaussian}} &= \Phi\left(\frac{\mu(\usermax)}{2} - \frac{\epsilon}{\mu(\usermax)}\right) - e^{\epsilon} \Phi\left(-\frac{\mu(\usermax)}{2} - \frac{\epsilon}{\mu(\usermax)}\right) \nonumber\\
\delta_{\text{infinite}} &= 1 - \Phi\left(\frac{\tauhi-\taulow}{\sigma}\right)^{\usermax}.
\end{align}
Further, for $\epsilon \geq 0$ as assumed here, $0 \leq \delta_{\text{Gaussian}} < 1$.

\subsubsection{Lower-bound derivation}
The lower-bound on $\delta$ follows from the first term in the three-term maximization of Equation~\ref{eq:final_result} and the third term in the three-term maximization evaluated at $\aplus = \usermax$ and $\aequals = 0$.  The first term is identically $\delta_{\text{infinite}}$ and the third gives $\delta_{\text{Gaussian}}$.

When $\usermax=1$, both the second and third terms are equal to $\delta_{\text{Gaussian}}$ because they are optimizations over $\aplus > 0$ so they can only be evaluated at $\aplus = 1 = \usermax$ and $\aequals=0$.  Hence the minimal $\delta = \max(\delta_{\text{infinite}}, \delta_{\text{Gaussian}})$ when $\usermax=1$.

\subsubsection{Upper-bound derivation}

We start by recalling $f$ from Eq.~\ref{eq:f} and that the partial derivative of $f$ with respect to $\mu$ is always positive, and that the partial derivative with respect to $\epsilon$ is always negative.  

To simplify, let function $\beta(\aequals) = \Phi\left(\frac{\tau^{*} - \tau}{\sigma}\right)^{\aequals}$. Then
the second and third terms of the three-term maximization in Theorem~\ref{th:gshm_privacy} written in terms of $\beta$ and $f$ are:
\begin{align}
&\max_{a_{+} + a_{=} \leq C_u, a_{+} > 0} 1 - \beta + \beta f(\mu(a_{+}), \epsilon - \ln \beta), \nonumber\\
&\max_{a_{+} + a_{=} \leq C_u, a_{+} > 0} f(\mu(a_{+}), \epsilon + \ln \beta)
\end{align}
Then $\mu(\aplus)$ is maximized when $\aplus = \usermax$.  Let $\mu^{*} = \mu(\usermax)$.  Given that the partial derivative of $f$ with respect to $\mu$ is always positive and the partial derivative with respect to $\epsilon$ is always negative (and $\ln \beta < 0$), we can write upper bounds for both terms as
\begin{align}
&\max_{a_{=} \leq C_u-1} 1 - \beta + \beta f(\mu^{*}, \epsilon), \nonumber\\
&\max_{a_{=} \leq C_u-1} f(\mu^{*}, \epsilon + \ln \beta)
\end{align}
Because $f(\mu^{*}, \epsilon) = \delta_{\text{Gaussian}} < 1$, the solution to the first optimization is $\aequals = \usermax-1$, which evaluates to a quantity even larger when $\aequals = \usermax$.  Evaluated at $\usermax$, the first equation is $\delta_{\text{infinite}} + (1-\delta_{\text{infinite}})\delta_{\text{Gaussian}}< \delta_{\text{infinite}} + \delta_{\text{Gaussian}}$, our desired upper-bound.  So what remains to be shown is that the second equation is less than or equal to the first.

Define
\begin{align}
r(\beta) &= 1 - \beta + \beta f(\mu^{*}, \epsilon), \nonumber\\
t(\beta) &= f(\mu^{*}, \epsilon + \ln \beta)
\end{align}
We will show $r(\beta) \geq t(\beta)$ for continuous $\beta \in [0, 1]$, the relevant range of $\beta$ for the above maximization over $\aequals$.  First, we note equality at the endpoints $r(0) = t(0) = 1$ and $r(1) = t(1) = \delta_{\text{infinite}}$.  Then $\frac{dr}{d\beta} = -1 + \delta_{\text{infinite}}$ is a constant negative slope and $\frac{dt}{d\beta} = \frac{\partial f(\mu^{*}, \epsilon + \ln \beta)}{\partial \epsilon} \frac{1}{\beta} = -e^{\epsilon} \Phi\left(\frac{-\mu^{*}}{2} - \frac{\epsilon + \ln \beta}{\mu^{*}}\right)$.  Evaluated at $\beta = 0$, $\frac{dt}{d\beta} < \frac{dr}{d\beta}$ because $-e^{\epsilon} < -1 + \delta_{\text{infinite}}$.  So slightly above $\beta = 0$, we have that $r > t$.  Remembering that $e^y \phi(\frac{x}{2} + \frac{y}{x}) = \phi(\frac{x}{2} - \frac{y}{x})$, where $\phi$ is the pdf for the standard normal, we have that
\begin{align}
\frac{d^2 t}{d \beta^2} = \frac{\phi\left(\frac{\mu^{*}}{2} - \frac{\epsilon + \ln \beta}{\mu^{*}}\right)}{\beta^2 \mu^{*}}.
\end{align}
This is always positive on the range of $\beta \in (0, 1]$.  

Now we claim via the mean-value theorem applied to the difference of the functions $r-t$, that because $\frac{d^2 (r-t)}{d \beta^2} < 0$ over $\beta \in (0, 1]$, and the two functions are equal at $\beta=0$ and $\beta=1$, there can be no other value of $\beta$ such that $r = t$ over this range. Since $r-t > 0$ slightly above $\beta = 0$, therefore $r(\beta) \geq t(\beta)$ for $\beta \in [0, 1]$, and we have proven our upper-bound.

\subsection{Corollary~\ref{cor:ratio_threshold}}

For $\usermax=1$ and $\delta \geq \delta_{\text{Gaussian}}$, ``add the deltas'' accounting would requires that 
\begin{align}
    \delta_{\text{infinite}} = \delta - \delta_{\text{Gaussian}} \geq 1-\Phi\left(\frac{\tauhi-\taulow}{\sigma}\right).
\end{align}
Exact accounting would require that
\begin{align}
    \delta \geq 1-\Phi\left(\frac{\tauhi-\taulow}{\sigma}\right).
\end{align}
Solving for the minimal $\tauhi-\taulow$ under the two cases (equality in the two expressions) and dividing gives the corollary.

\end{document}